\documentclass{article}
\usepackage{spconf,amsmath,graphicx,hyperref,amssymb, float}

\newcommand*{\email}[1]{%
    \normalsize\href{mailto:#1}{\texttt{#1}}\par
    }

\title{3D CBCT CHALLENGE 2024: IMPROVED CONE BEAM CT RECONSTRUCTION USING SWINIR-BASED SINOGRAM AND IMAGE ENHANCEMENT}
\name{Sasidhar Alavala, Subrahmanyam Gorthi}
\address{Department of Electrical Engineering, IIT Tirupati, India \\ \email{ansr2510@gmail.com}, \email{s.gorthi@iittp.ac.in}}

\begin{document}
\ninept
\maketitle
\begin{abstract}
In this paper, we present our approach to the 3D CBCT Challenge 2024, a part of ICASSP SP Grand Challenges 2024. Improvement in Cone Beam Computed Tomography (CBCT) reconstruction has been achieved by integrating Swin Image Restoration (SwinIR) based sinogram and image enhancement modules. The proposed methodology uses Nesterov Accelerated Gradient Descent (NAG) to solve the least squares (NAG-LS) problem in CT image reconstruction. The integration of sinogram and image enhancement modules aims to enhance image clarity and preserve fine details, offering a promising solution for both low dose and clinical dose CBCT reconstruction. The averaged mean squared error (MSE) over the validation dataset has decreased significantly, in the case of low dose by one-fifth and clinical dose by one-tenth. Our solution is one of the top 5 approaches in this challenge.
\end{abstract}
\begin{keywords}
Cone Beam Computed Tomography Reconstruction, Nesterov Accelerated Gradient Descent, Sinogram Enhancement, Image Enhancement, SwinIR
\end{keywords}
\section{INTRODUCTION}
\label{sec:introduction}

In medical imaging, particularly in CT, reducing the radiation dose is essential for patient safety. However, this reduction poses a significant diagnostic problem — the compromised quality of images due to lower signal-to-noise ratio and diminished information. Therefore, the problem demands solutions that balance diagnostic efficacy and the potential risks linked to excessive radiation exposure.

SwinIR, or Swin Image Restoration, introduced by J Liang et al. \cite{liang2021swinir} in 2021, deviates from conventional convolutional neural networks by adopting a hierarchical and shift-based design. Unlike traditional methods, SwinIR divides images into non-overlapping patches, processing them through successive stages, and allowing for more efficient feature extraction. 

There is continuous development in the field of CBCT for better reconstruction algorithms than the standard FDK algorithm. Even though there are some ML-based algorithms, iterative algorithms are still considered best. However, iterative algorithms take longer computation time. So, in our approach, we have used NAG to accelerate the convergence.

\section{METHODOLOGY}
\label{sec:method}

\begin{figure}[htb]

\begin{minipage}[b]{1.0\linewidth}
  \centering
  \centerline{\includegraphics[width=8.5cm]{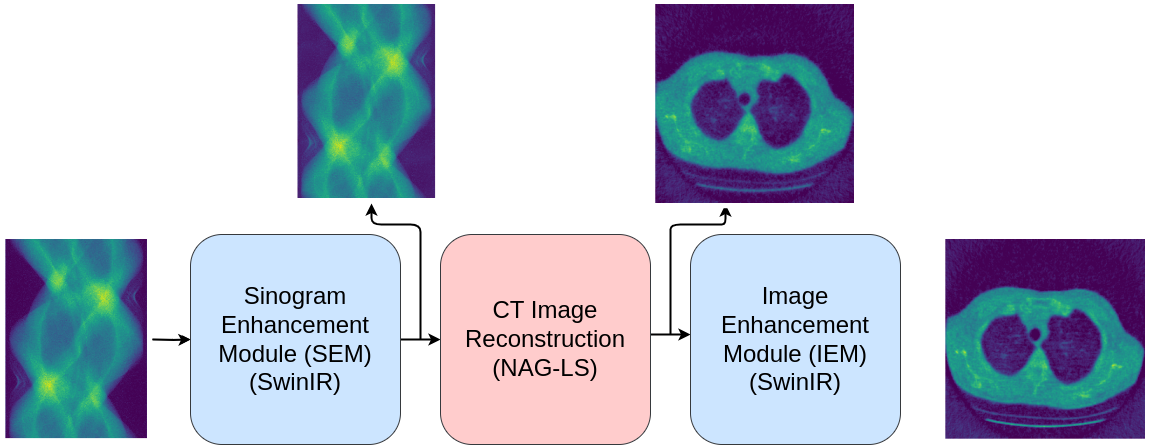}}
  \vspace{-0.5cm}
\end{minipage}
\caption{Proposed CBCT reconstruction pipeline.}
\label{fig:fig1}
\end{figure}

Our proposed method for improving CBCT reconstruction involves an integration of SwinIR-based sinogram and image enhancement modules (SEM and IEM), coupled with the application of NAG to solve the least squares problem in CT image reconstruction. SwinIR is employed to enhance the quality of both sinogram data and the reconstructed CBCT image. First, the low dose sinogram is given as input to SEM. By leveraging SwinIR, we effectively capture fine details and reduce noise in the sinogram, ensuring a more accurate representation of the underlying structures. This enhanced sinogram is then given as input to the CT image reconstruction module, where the least squares problem is iteratively solved to reconstruct the CBCT image. The objective function for the least squares problem in the context of CT image reconstruction is given by:
\[ \min_{x \in \mathbb{R}^{256 \times 256 \times 256}} f(x) = \frac{1}{2} \biggl\lVert Ax - b \biggr\rVert_2^2 \]

Here, \(A\) is the system matrix (forward operator) representing the CT geometry, \(x\) is the vectorized form of the reconstructed image, and \(b\) is the measured sinogram data. NAG \cite{nesterov1983method}, known for its accelerated convergence properties, efficiently guides the optimization process. Finally, the reconstructed CT image is given as input to IEM for enhancing CT image quality by preserving fine details.

\begin{figure*}[]

% \begin{minipage}[b]{1.0\linewidth}
  \centering
  \centerline{\includegraphics[width=15cm]{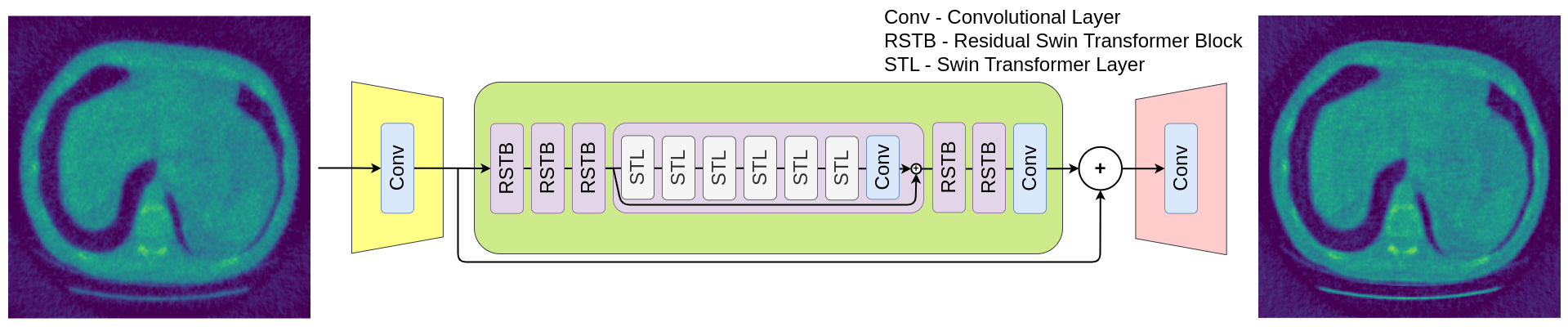}}
  \vspace{-0.5cm}
% \end{minipage}
\caption{Block diagram of SwinIR architecture \cite{liang2021swinir}.}
\label{fig:fig3}
\end{figure*}

\subsection{Training Details}
\label{ssec:training}

SwinIR \cite{liang2021swinir} architecture comprises three modules:

\begin{itemize}
\item Shallow feature extraction module: It consists of a convolution layer to extract the low-level features.
\item Deep feature extraction module: It consists of a series of residual swin transformer blocks (RSTB) followed by a convolution layer. Each RSTB comprises swin transformer layers with a residual connection. It extracts the high-level features which aren't captured in the earlier module.
\item Image reconstruction module: It consists of a convolution layer for reconstructing the CT image by aggregating both the output of shallow and deep feature extraction modules.
\end{itemize}

The SEM takes an input size of (256, 256) with 360 channels, reflecting the sinogram's views, while the IEM takes an input size of (256, 256) with 256 channels. The SEM is followed by a reconstruction module which takes an input sinogram of size (256, 256) with 360 channels and gives an output CT image of size (256, 256) with 256 channels, which employs NAG-LS from Tomosipo \cite{hendriksen2021tomosipo} to obtain the CT image. Key hyperparameters, including the learning rate, batch size, and regularization parameters, are systematically varied, with the learning rate set to 0.0005 for the Adam optimizer. The training for the two modules is done separately for 250 epochs each, and the MSE loss criterion guides the model training.

\section{RESULTS}
\label{sec:results}

The reported MSE values represent the average performance on the validation dataset for different CBCT reconstruction methods. Compared to FDK \cite{feldkamp1984practical} and the Simultaneous iterative reconstruction technique (SIRT), the NAG-LS method demonstrates significant improvements, achieving lower MSE values for low-dose CBCT.

\begin{table}[h]
\centering
\caption{Average MSE for validation set.}
\label{tab:table1}
\setlength{\tabcolsep}{7pt} % Adjust cell padding
\renewcommand{\arraystretch}{1.3} % Adjust row height

\begin{tabular}{|c|c|c|}
\hline
\textbf{Reconstruction Method} & \textbf{Low dose} & \textbf{Clinical dose} \\
\hline
\textbf{FDK (Baseline)} & 0.07959  & 0.03102 \\
\hline
\textbf{SIRT} & 0.06648 & 0.04545 \\
\hline
\textbf{NAG-LS} & 0.04408 & 0.04070 \\
\hline
\textbf{NAG-LS with SEM} & 0.01520 & 0.00940 \\
\hline
\textbf{NAG-LS with SEM and IEM} & \textbf{0.00918} & \textbf{0.00467} \\
\hline
\end{tabular}
\end{table}

% Below is an example of how to insert images. Delete the ``\vspace'' line,
% uncomment the preceding line ``\centerline...'' and replace ``imageX.ps''
% with a suitable PostScript file name.
% -------------------------------------------------------------------------
\begin{figure}[htb]

\begin{minipage}[b]{1.0\linewidth}
  \centering
  \centerline{\includegraphics[width=8.5cm]{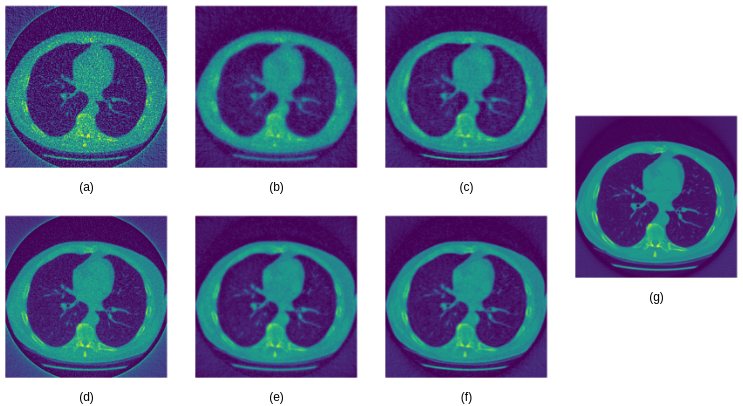}}
  \vspace{-0.5cm}
\end{minipage}
\caption{(a), (d) are CT images reconstructed using FDK (Baseline). (b), (e) are CT images reconstructed using NAG-LS with SEM. (c), (f) are CT images reconstructed using NAG-LS with SEM and IEM. Upper and lower row images are low dose and clinical dose respectively. (g) is the clean CT image.}
\label{fig:fig4}
\end{figure}

However, for clinical dose scenarios, FDK outperforms NAG-LS, showcasing lower MSE values (0.03102 compared to 0.04070) but this performance is not continued when SEM is used where NAG-LS outperformed FDK again. The most substantial decrease in error is observed with the combined approach of NAG-LS, SEM, and IEM, yielding the lowest MSE values of 0.00918 for low-dose and 0.00467 for clinical-dose reconstructions on the validation set. The average MSE values on the hidden test set are similar to the validation set, 0.008837 for low-dose and 0.004480 for clinical-dose. Despite the occurrence of considerable blurring, the IEM has effectively mitigated this effect by enhancing the image, accentuating sharp features and making them more prominent.

\section{CONCLUSION}
\label{sec:conclusion}

In conclusion, our approach for this 3D CBCT Challenge 2024 \cite{Biguri_Mukherjee_2023} has significantly reduced MSE which involves enhancing the sinogram first using SEM and then reconstructing the CT image from it by NAG-LS approach iteratively and then again enhancing the CT image using IEM for fine detail preservation.

\section{ACKNOWLEDGEMENT}
\label{sec:ack}
This research has received partial support through project grant number 2022-IRP-26694272 from the Semiconductor Research Corporation (SRC) under the India Research Program (IRP).

% References should be produced using the Bibtex program from suitable
% BibTeX files (here: strings, refs, manuals). The bib. best bibliography
% style file from IEEE produces an unsorted bibliography list.
% -------------------------------------------------------------------------
\bibliographystyle{IEEEbib}
\bibliography{refs}

\begin{thebibliography}{1}

\bibitem{liang2021swinir}
Jingyun Liang, Jiezhang Cao, Guolei Sun, Kai Zhang, Luc Van~Gool, and Radu Timofte,
\newblock ``Swin{IR}: Image restoration using swin transformer,''
\newblock in {\em Proceedings of the IEEE/CVF international conference on computer vision}, 2021, pp. 1833--1844.

\bibitem{nesterov1983method}
Yurii~Evgen'evich Nesterov,
\newblock ``A method of solving a convex programming problem with convergence rate \(\mathcal{O}\left(\frac{1}{k^2}\right)\),''
\newblock in {\em Doklady Akademii Nauk}. Russian Academy of Sciences, 1983, vol. 269-3, pp. 543--547.

\bibitem{hendriksen2021tomosipo}
Allard~A Hendriksen, Dirk Schut, Willem~Jan Palenstijn, Nicola Vigan{\'o}, Jisoo Kim, Dani{\"e}l~M Pelt, Tristan Van~Leeuwen, and K~Joost Batenburg,
\newblock ``Tomosipo: fast, flexible, and convenient 3d tomography for complex scanning geometries in python,''
\newblock {\em Optics Express}, vol. 29, no. 24, pp. 40494--40513, 2021.

\bibitem{feldkamp1984practical}
Lee~A Feldkamp, Lloyd~C Davis, and James~W Kress,
\newblock ``Practical cone-beam algorithm,''
\newblock {\em Josa a}, vol. 1, no. 6, pp. 612--619, 1984.

\bibitem{Biguri_Mukherjee_2023}
Ander Biguri and Subhadip Mukherjee,
\newblock ``Advancing the frontiers of deep learning for low-dose 3{D} cone-beam computed tomography ({CT}) reconstruction,''
\newblock in {\em IEEE International Conference on Acoustics, Speech and Signal Processing (ICASSP)}. IEEE, 2024.

\end{thebibliography}

\end{document}